\documentclass[8pt]{article}
\usepackage[export]{adjustbox}
\usepackage{amsthm}

\usepackage{nccmath}
\usepackage{mathrsfs}
\usepackage{balance}
\usepackage{etoolbox}

\usepackage[utf8]{inputenc}
\usepackage{flushend}
\usepackage[section]{placeins}
\usepackage[bf, footnotesize]{caption}
\usepackage{graphics}
\usepackage{CJKutf8}

\usepackage{amssymb}
\usepackage{amsthm}
\usepackage{mathtools, cuted}
\usepackage{widetext}

\usepackage{lineno}
\usepackage{array}
\usepackage{hyperref}
\usepackage{multicol}
\usepackage{booktabs}
\usepackage{amsmath}
\usepackage{widetext}
\usepackage{flushend}
\usepackage[explicit]{titlesec}
\usepackage{authblk}

\titleformat{\section}
  {\normalfont}{\thesection}{1em}{\MakeUppercase{#1}}
  \titleformat{\subsection}
  {\it}{\thesection}{1em}{{#1}}

\begin{document}
\small

\title{Analysis of how a consciousness-lacking mechanistic observer perceives real events and those simulated by a computer}
\date{}


\author[1]{Francesco Sisini}

\affil[1]{Department of Applied Research Tekamed Ltd. Via Bellaria 6, 44121 Ferrara ,Italy}



\begin{center}

\end{center}
\maketitle

\section*{Keywords}
Simulation, virtual reality, inertial observer, Oculus, physical computation

\section{Abstract}

Physical theories must stem from observation. The possibility that perceived events are simulated, not real, raises a crucial dilemma about the credibility of known physics, known as the simulation hypothesis. To analyze this hypothesis in deterministic and mechanistic terms, an observer is conceptually developed, whose internal state depends on external events. It is demonstrated that the observer cannot distinguish between simulated and real events. It is argued that if the observer could observe its entire dynamics, it could formulate the necessary principles of relativity to distinguish real from non-real physical motions. Finally, an attempt is made to apply the same considerations developed for the mechanistic observer to the human being.

%
%

\section{Introduction}

It is known that at the foundation of currently accepted physical theories lies a significant epistemological problem, namely, the uncertainty about whether the physical laws underlying these theories were deduced from observing natural phenomena or artificially produced simulations. Though seemingly counterfactual, this hypothesis has intrigued minds from ancient times to the present day \cite{bostrom,enwiki:1170540473,enwiki:1189457934}. This article examines this hypothesis by conceptually constructing a consciousness-devoid mechanistic observer (MO), whose internal state is entirely determined by classical physical laws, perceiving classical physical events through an artificial device analogous to human eyes. The analysis is confined to classical, non-relativistic systems.

\section{Mechanistic Observer}
The observer that will be described below is not innovative, and it could even be avoided to delve into its technological details. Nevertheless, I believe it is useful to provide a brief description of its implementation to follow the discourse developed in the article and envision what is hypothesized later on.\\
The considered MO consists of a pair of photodetectors placed at a distance $d$ from each other. Event detection is contingent on the presence of light in the assumed environment. Each detector has an arbitrary aperture diaphragm and is composed of a square matrix of photosensitive elements. An event $(t,x,y,z)$ is detected by both photodetectors. The coordinates $x, y, z$ of the event can be derived from the event's detection coordinates on the first (detector 1) and second (detector 2), namely $x_1, y_1$ and $x_2, y_2$, utilizing parallax.
At time $t$, the event detection determines the state $\psi(t)=(x_1(t),y_1(t),x_2(t),y_2(t))$ of the MO. If we denote $\mathbf{x_i}: I\rightarrow \mathbb{R}^{3}$ as a differentiable mapping from the interval $I$ of the real axis to $\mathbb{R}^{3}$, this mapping is called the motion of the $i$-th material point in $R^3$. If only the material point labeled with $i$ is present in the MO's field of view, then the motion $\mathbf{x_i}$ completely determines the evolution of the MO's state $\psi$.

In general, given a system of $N$ material points visible to the MO, their global motion $\mathbf{x}:I\rightarrow \mathbb{R}^{3N}$ determines the temporal evolution of the MO's state $\psi$. The MO may also be equipped with a system that periodically records its internal state and a computing automaton capable of comparing two different recordings.

\section{Equivalent Motions and Simulations}

Consider two labeled systems, denoted as system $a$ and system $b$, both comprising $N$ material points with respective motions $\mathbf{x_{(a)}}$ and $\mathbf{x_{(b)}}$. Let's consider an equivalence relation called \textit{equivalence}, where the two motions are equivalent if there exist a  Galilean transformations $g$ such that $\mathbf{x_{(a)}}=g \mathbf{x_{(b)}}$. In essence, two systems are equivalent if there exist two inertial reference frames in which they correspond instant by instant. It's interesting to note that equivalence between motions can be recognized by an MO, which, without additional information, couldn't distinguish one motion from another.

Alongside this relation, the concept of simulation between two motions is introduced as a less stringent equivalence. The idea is to identify a combination of $k$ coordinates of material points in system $a$ such that their motion for a certain time interval and under certain initial conditions coincides with a combination of $k$ coordinates of system $b$. For example, considering system $a$ with two masses $m_{(a),1}$ and $m_{(a),2}$ subject to a quadratic potential and system $b$ with two masses $m_{(b),1}$ and $m_{(b),2}$ subject to a Keplerian potential, it can be easily verified that there exist $x_{(a),i}$ and $x_{(a),j}$ in system $a$ and $x_{(b),l}$ and $x_{(b),m}$ in system $b$ such that, with an appropriate choice of initial conditions, $x_{(a),i}(t) = x_{(b),l}(t)$ and $x_{(a),j}(t) = x_{(b),m}(t)$ for a certain time interval.

We define two systems as a simulation of each other if there exist initial conditions for the coordinates of both systems and an interval $\Delta I$ such that a combination $C_k$ of coordinates of $a$ is equivalent to a combination $C'_k$ of coordinates of $b$. The concept of simulation can be made more stringent by requiring that this relationship is maintained between the two systems not for a precise choice of initial conditions, but at least for some neighborhood of them. We define this type of simulation as \textit{strong simulation}. As it is easy to verify, for two systems to be in a strong simulation relationship, they must differ only in the values of the parameters that characterize them (Coulomb force and gravitational force), or one of the two systems must contain the other as an isolated subsystem. For example, a system $a$ consisting of a harmonic oscillator and a Keplerian system that do not interact with each other can be equivalent to a system $b$ consisting of a harmonic oscillator. It can be asserted that, given a system $a$, there is always a \textit{weak} simulation, but there is no strong simulation that does not fall into the two trivial cases described above.

From the above, it follows that an MO cannot distinguish between two equivalent motions, but two motions can be equivalent only if the underlying physical systems correspond to each other. If they do not correspond, their equivalence is limited to a specific choice of initial conditions for the systems. This aspect is relevant because it can be inferred that an MO capable of interacting with systems could distinguish them by observing how their motions change with varying initial conditions.

\section{Computer Simulations of Motions}

The result obtained in the previous paragraph appears to be at odds with the common experience of computer systems simulating any physical motion. Computers are, in every respect, physical systems; therefore, even computers cannot simulate other physical systems strongly. For instance, it is known that the physics describing internal motions within a computer is not the same as that used to describe a Keplerian motion, or more simply, there are no Keplerian motions in a computer.

To reconcile the definition given here of simulation with computer science, we introduce the concept of super coordinates. Here, a combination $C_m$ of $m$ coordinates $x_i$ (of a physical system) is defined as a super coordinate $\mathscr{X}$ of the physical system with respect to a certain function $f$, expressed as the relation $\mathscr{X} = f(C_m)$. A common example of a super coordinate is the function $f(x_0,...,x_7)=x_0\times2^0+...x_7\times 2^7$, which associates an integer from 0 to 255 with the charge of eight capacitors. If the simulation relationship is applied between the coordinates of a system $a$ and the super coordinates of a system $b$, then it can be a strong simulation, as evident from common experience.

In conclusion, it can be asserted that the super coordinates of a computer can evolve over time to be in a strong simulation relationship with any (classical) physical system. Taking a generic system $a$ and the computer simulating it ($b$), there is still a distinction between the physical system $a$ which can be considered real, and the physical system $b$ described by the super coordinates of the computer. The super coordinates are not Cartesian coordinates of real entities; rather, they can be considered as noumena, as they conceptually represent the coordinates but are not the coordinates themselves. The difference between the motion of system $a$ and system $b$ becomes evident when analyzing the two systems through the MO. For the MO, the temporal evolution of the state $\psi_a$ induced by the motion $\mathbf{x_{(a)}}$ cannot be directly equated, instant by instant, to the state $\psi_b$ induced in the MO by observing the computer, even if every internal mechanism of the computer were exposed to the MO's view. Therefore, the computer simulation of a physical system $a$ is not considered such by an MO, which, on the contrary, will find the two systems distinct and not bound by a simulation relationship.
\section{Virtual Reality}

This initial result might seem discouraging for the hypothesis of simulating reality. Indeed, establishing that an automaton like the MO can distinguish computer simulation from reality might appear to be the premise for formulating a no-go theorem. However, attempting to formulate such a theorem reveals a first loophole. Consider an optical device connected to computer $b$ to be applied to the MO after removing the diaphragm in front of the detectors. Such a device can be easily designed to create optical stimuli corresponding to the values taken by the super coordinates of the connected computer.
Therefore, using such a device, a computer simulation for which $\mathbf{x}{(a)}=g \mathbf{\mathscr{X}}{(b)}$ induces the same state in the MO as that induced by the simulated physical system, deceiving the MO, which cannot distinguish between reality and simulation.

This positive result toward the rationality of the simulation hypothesis raises the question of whether an MO can at least distinguish between a physical motion and one generated arbitrarily through an algorithm, such as the one hypothesized to place the concept of free will in a deterministic context \cite{hossenfelder2012free}. To formulate this question correctly, it will be assumed later on that the MO is connected to a universal computer, enabling it to process its internal state.

\section{Autogenous Epistemological Principle for the Mechanistic Observer}
The problem of distinguishability between a physical motion and an arbitrary one becomes interesting when the conceptual tools of the MO are appropriately limited. Indeed, it is evident that if we consider the MO to be already trained to distinguish these two possibilities using the known physical principles, it can make this distinction automatically.\\
Therefore, it is assumed that the only point of reference available to the MO is a principle referred to here as the \textit{autogenous} principle, according to which a motion is physical if it can be reproduced by the MO itself through a suitable algorithm, subject to the condition that the algorithm's complexity is not maximum. In other words, the algorithm should not be a mere listing of coordinates but should involve a certain computational procedure. With only this principle at its disposal, the MO will classify as physical any motion that does not correspond to free will. However, among the motions classified as physical, there will also be non-physical motions, such as the acceleration of an isolated point particle. Therefore, using this principle, the MO will classify every algorithmically generated motion by a computer as physical.\\
This result informs us that an MO properly equipped with the mentioned viewer not only can be deceived about the \textit{real} nature of perceived events but also lacks a native tool to distinguish what could be real from what could not even in principle. Obviously, this latter consideration imposes limits on the ability of an MO to formulate a consistent physical theory even when equipped with machine learning and an adequate observation system.
\section{From Observing External Events to Internal Ones of the Mechanistic Observer}

The result obtained in the previous paragraph seems to lean toward a discouraging hypothesis regarding physical realism. Limiting humans to perceptual and rational capacities, one might conclude that we, too, lack an autogenous mechanism to distinguish physical motions from arbitrarily generated ones. This impasse can be overcome if the MO has the ability to observe its internal motions without the internal view being controlled by the viewer. From these motions, the MO can algorithmically derive, through machine learning, the principles of relativity and then use them as a criterion to distinguish physical from non-physical motions.
\section{Discussion}

Mutatis mutandis, the argumentation put forth for the MO can be used to conduct a similar analysis regarding the possibility that humans can distinguish whether the perceived reality is real or a result of simulation.

Humans, equipped with an appropriate system to interface with physical stimuli from the real world, could be the subject of a simulated reality. Although this hypothesis has been examined in this article only limitedly in the context of a classical universe, it has also been investigated in the quantum and relativistic context \cite{Chung2016WeAL}

Based solely on the autogenous principle, "anything that can be described by a formula is physical" the boundary between mathematics and physics would be eliminated. This could lead to considering even non-physical motions as real, such as an isolated particle following a non-linear and uniform motion, and thus having no way to escape, effectively being unable to distinguish reality from simulation.
Similarly to what was seen for the MO, humans could, however, understand the general principles of physics through observing the mechanisms governing their mental activity rather than observing the universe.

From these principles, they could derive universally valid criteria to assess the reality of their own perceptions. However, this scenario seems to require that humans have a consciousness capable of clearly distinguishing themselves from the universe.\\
The presence of consciousness in the human being is somewhat accepted without reservations, as there are currently no rational criteria to establish the presence or absence of consciousness in a physical system, including humans and computers. The interest of neurophysiologists like Giorgio Vallortigara is increasingly focused on studying the role played by consciousness in intelligence, while the attention of theoretical physicists is gradually shifting from attempting to unify the treatment of fundamental interactions to interpreting natural phenomena in terms of computational processes\cite{whitworth,muller,marletto,horsman17,Dehghani}.

This new paradigm, called info-computational, takes information and computational processes as fundamental ontological entities rather than space-time and matter, and the laws of nature reflect relationships between informational structures rather than the coordinates of material points\cite{muller}.
The autogenic principle presented here, observed from the info-computational perspective, could be functional in investigating the laws of the universe, as the computational principles of the brain might be similar to those of the universe. In this scenario, the autogenic principle could become \textit{si possum cogitare, ergo potest esse} (if I can think it, then it can exist), paraphrasing Descartes.
In the info-computational view, the simulation hypothesis loses its intrinsic interest as it is effectively absorbed into the very principles of the physical theory because it does not require the intervention of a metaphysical will to be implemented.
It is my opinion that this paradigm will dominate research in the coming decades, but before fully embracing it, I believe it is appropriate to delve into how observing physical events through the senses aligns with observing internal physical events within the observer, which could be done using one's consciousness.

\section{Conclusions}

A mechanistic observer, with the sole means to distinguish physical systems being the comparison of their motions, may be induced to confuse a real physical motion with one simulated by a computer system. If such a system lacks foreknowledge of the principles of Galilean relativity, it cannot develop, through pure observation of external events, a criterion to distinguish physical motions from non-physical ones that might be presented to it. If the observer can also observe events internal to itself, it can use that experience to derive the principles of relativity necessary to distinguish physical from non-physical motions.

\bibliographystyle{unsrt} 
\bibliography{siso}

\end{document}